\documentstyle[epsf]{elsart}
\begin{document}
\begin{frontmatter}
\newcommand{\postscript}[2]{\setlength{\epsfxsize}{#2\hsize}
\centerline{\epsfbox{#1}}}
\title{Renormalization of the one-pion-exchange interaction}
\author[ITA]{T. Frederico,}
\author[USP]{V.S. Tim\'oteo,}
\author[IFT]{Lauro Tomio}
\address[ITA]
{Departamento de F\'\i sica, Instituto Tecnol\'ogico de
Aeron\'autica, CTA, \\
12228-900, S\~ao Jos\'e dos Campos, Brasil}
\address[USP]{Nuclear Theory and Elementary Particle Phenomenology
Group,\\
Instituto de F\'{\i}sica, Universidade de S\~{a}o Paulo,
05315-970, S\~{a}o Paulo, Brasil}
\address[IFT]{Instituto de F\'\i sica Te\'orica, Universidade Estadual
Paulista,  01405-900, S\~{a}o Paulo, Brasil}
\date{\today} 
\maketitle 
\begin{abstract}
A renormalization scheme for the nucleon-nucleon (NN) interaction based 
on a subtracted T-matrix equation is proposed and applied to the 
one-pion-exchange potential supplemented by contact interactions. 
The  singlet and triplet scattering lengths are given to fix the 
renormalized strengths of the contact interactions. 
With only one scaling parameter ($\mu$), the results show an overall very
good agreement with neutron-proton data, particularly for the
observables related to the triplet channel. 
The agreement is qualitative in the $^1S_0$ channel.
Between the low-energy NN observables we have examined, the mixing
parameter of the $^3S_1-^3D_1$ states is the most
sensible to the scale. The scheme is renormalization group invariant
for $\mu\to\infty$.
\end{abstract} 
\end{frontmatter}

\section{Introduction}
The effective field theory (EFT) of nuclear forces based on a chiral
expansion of an effective Lagrangian, as proposed by Weinberg 
\cite{wei}, gives a nucleon-nucleon (NN) interaction in
leading order which is the one-pion-exchange potential (OPEP) plus a 
Dirac-delta.
He suggested to infer the values of the strength of the delta force in the
$^1S_0$ and $^3S_1$ channels from the singlet and triplet scattering
lengths respectively.
 Also such potential should be valid for momenta well below a
typical momentum scale of quantum chromodynamics such as the rho meson mass 
($m_\rho \sim$ 4 fm$^{-1}$)~\cite{wei}. Therefore, the 
intermediate virtual propagation of the NN system should be cut at the
momentum scale of this order.

In spite of the significative results obtained by several authors, when 
implementing the EFT program for the two nucleon 
system~\cite{lepage,bira,kaplan,meissner,birse,rho,mehen}, 
and also by a recent related approach~\cite{perry}, as far as we know the
predictibility power of the leading order term (OPEP plus delta) with a 
single renormalization  momentum scale was not fully explored in the 
literature.\ The result of such calculation would give a sound basis for
the renormalization program of EFT in NN system. 
The leading order term was already considered in NN calculations, 
when comparing with more realistic calculations~\cite{BR}, or as 
part of interactions used in the EFT 
program~\cite{bira,kaplan,cohen,gegelia}. \ 
However, we argue that, without a systematic analysis of the physical
contribution coming from each order term in the renormalization 
procedure, it is easy to miss their real significance when parametrizing
the full interaction to obtain the desired observables. \ 
To start, it is relevant to know how much physical
information can one extract from the leading order term, using a
renormalization procedure. What are the observables one can better
describe with just the leading order term (the higher order terms should
not affect too much these observables) and what are the observables for
which higher order terms are essential? Our aim in this paper is to answer
such questions. We hope the following analysis can be useful in the 
renormalization program of the effective field theory applied to 
the nucleon-nucleon system. 

In this work we have verified strictly the validity of Weinberg 
suggestion, to first consider (in leading order) an effective potential
which ``consists of just a conventional one-pion exchange term, plus a 
direct neutron-proton ($n-p$) interaction produced by the four-fermion
terms"~\cite{wei}. \ Such ``four-fermion terms"  consist of two delta-function
interactions: one in the $^1S_0$ channel and the other in the $^3S_1$
channel.  \ To be consistent with the original suggestion, 
to cut the virtual propagation of the NN system at the scale $m_\rho$, 
we first use a renormalization scale of such order. \ 
However, we understand this as a qualitative suggestion,
 such that we explore its limitation in the calculation of several 
low-energy NN observables. So, we finally consider a free renormalization 
scale in our numerical calculation, and try to obtain a consistent
description, in order that such scale vanishes
as a physical parameter as required by a renormalization group invariant 
theory.

In our approach, we use a renormalization scheme for the T-matrix, 
which is appropriate when considering the OPEP plus a
delta potential. It incorporates the momentum scale $(\mu)$ where the 
two-nucleon propagation is removed and  the scattering lenghts  fix the
renormalized strengths. \ A subtracted T-matrix integral equation is derived, 
where the subtraction is done at energy of $-\mu^2$, in the center
of mass system (in units of $\hbar=1$ and nucleon mass $m=1$). \ 
The input is the T-matrix at energy $-\mu^2$, which is chosen as the
OPEP supplemented by delta interactions. The scale $\mu$ is 
of the order of $m_\rho$, and we will show that the results for the 
renormalized T-matrix, in the limit of $\mu \rightarrow \infty$, once the 
scattering lengths are kept fixed, are quite consistent with 
$\mu\sim m_\rho$, at least for the diagonal components of the matrix
(the extension of this statement will be clarified).
As we verified, the numerical results in the $^3S_1-^3D_1$ channels,
obtained by following this procedure are in good qualitative 
agreement with the experimental data  for the $n-p$ phase-shifts, mixing
parameter, the deuteron binding energy and D/S ratio, at laboratory 
energies below $\sim$ 50 MeV. 

The higher order terms in the EFT expansion of the effective interaction
will become more important for laboratory energies above 50 MeV in the
$^3S_1-^3D_1$ channels, where the leading order results clearly 
deviate from the data. However, in the $^1S_0$ state the importance of
the higher order terms appears already in the value of the effective
range, which is underestimated in lowest order. 
Such results confirm that the effective interaction in the singlet 
state needs higher order correction (as already expected), but such
corrections should not affect in an essential way the triplet states
(particularly at low energies), which are already in a fair agreement
with data. \  

In our scheme, it is considered an unregulated effective
power expansion of the NN potential only in the mid- and short-
range parts of the interaction, such that the well stablished 
long-range part (OPEP) is kept intact~\footnote{The OPEP is common 
to all realistic interactions, and an effective power expansion which
does not consider it will have to face difficulties with the 
parametrization, as the constants should represent simultaneously
long and short range physics.}.
So, the matrix elements of the full effective NN interaction, 
in the relative momentum space, will be given by
\begin{eqnarray}
\langle\vec {p'}|V_{EFT}|\vec p \rangle
= \langle\vec {p'}|V_\pi|\vec p \rangle &+& \frac{1}{2\pi^2}
\left[
\frac{1 - \vec\tau_1\cdot\vec\tau_2}{4}
\left( \lambda_t^{(0)} + \lambda_t^{(1)}p^2
 + \lambda_t^{(1)*}{p'}^2 + \cdot \cdot \cdot \right) 
+ \right. \nonumber \\
&+& \left. \frac{3 +\vec\tau_1\cdot\vec\tau_2}{4}\left(\lambda_s^{(0)} +
\lambda_s^{(1)}p^2 + \lambda_s^{(1)*}{p'}^2 + \cdot \cdot \cdot \right) 
\right], \label{veft}
\end{eqnarray} 
where the unregulated strengths of the contact interaction, and the higher
order derivatives, have subindices $t$ and $s$, for the isospin singlet
(spin triplet) and isospin triplet (spin singlet) channels respectively.
The matrix element of the one-pion-exchange potential is given by:
\begin{eqnarray}
\langle\vec {p'}|V_\pi|\vec p \rangle
=-\frac{g_a^2}{4(2\pi)^3f_\pi^2} \vec\tau_1\cdot\vec\tau_2 
\frac{\vec\sigma_1\cdot(\vec {p'}-\vec {p})\;  
 \vec\sigma_2\cdot(\vec {p'}-\vec {p})}
{ (\vec{p'}-\vec{p})^2+m_\pi^2}\ ,
\label{opep}
\end{eqnarray} 
where $\sigma_i$ and $\tau_i$ are the usual spin and isosping Pauli
matrices for nucleon $i$; $g_a\;(=1.25)$ is the axial coupling constant,
$f_\pi \; (=93$ MeV) is the pion weak-decay constant, and 
$m_\pi\; (=138$ MeV) is the pion mass. 
${\vec p}^\prime- \vec p$ is the momentum transfer.
The leading order term of the expansion, which we are going to consider
in the next sections, imply to take $\lambda_{s(t)}^{(n)} = 0$ for all
$n\ge 1$ in eq.~(\ref{veft}). 

In section II we present the formalism for the subtracted T-matrix, 
which is applied to the leading order term of the NN interaction.
In section III we present our main results and conclusions.

\section{Subtracted T-matrix and the leading order NN
interaction}
In the next we introduce our formalism which is based on a subtracted
equation for the T-matrix.
We begin with the formal T-matrix equation (for a regular potential), which
is written as:
\begin{eqnarray}
T(E)&= V+VG^{(+)}_0(E)T(E) &= [1-VG^{(+)}_0(E)]^{-1}V
\label{tren1} \\
&= V+T(E)G^{(+)}_0(E)V &= V[1-G^{(+)}_0(E)V]^{-1} \ ,
\label{tren2}
\end{eqnarray} 
where the free Green's function for the two-body system, with
the appropriate boundary condition, in terms of the free 
Hamiltonian $H_0$, is
\begin{equation}
G^{(+)}_0(E)=(E+i\varepsilon -H_0)^{-1} .\label{G0}
\end{equation}

Considering that our input is given by the T-matrix at
a given scale, the potential will formally be replaced by the 
T-matrix at scale $\mu$ through Eq.(\ref{tren1}) or 
Eq.(\ref{tren2}):
\begin{eqnarray}
V=T(-\mu^2)\left[{1+G_0(-\mu^2)T(-\mu^2)}\right]^{-1}
=\left[{1+T(-\mu^2) G_0(-\mu^2)}\right]^{-1}T(-\mu^2)
\ . 
\label{vren}
\end{eqnarray}
Substituting the potential given by Eq.(\ref{vren}) in  Eqs. (\ref{tren1}) 
and (\ref{tren2}), the corresponding subtracted equation for the T-matrix 
is found:
\begin{eqnarray}
T(E)&=&T(-\mu^2)+T(-\mu^2)
\left( G^{(+)}_0(E)- G_0(-\mu^2)\right)T(E) 
\label{tren3}
\end{eqnarray} 
One should note that we have chosen a real value for $\mu$
(negative energy), but this is not a restriction to the generality 
of the subtracted T-matrix equation (\ref{tren3}).
However, such choice has the advantage that the kernel of the
integral equation has just one fixed point singularity for positive
energies $E$, and the same formal structure as the original T-matrix 
equation.  \ As one can see, from the original T-matrix, one  
simply needs to replace $V$ by $T(-\mu^2)$ and multiply the
free propagator by a function which depends on the scale $\mu$, 
such that 
\begin{eqnarray}
T(E)&=&
T(-\mu^2)+T(-\mu^2)G^{(+)}_R(E;-\mu^2)T(E)
\label{tr1}
\end{eqnarray}
where
\begin{equation}
G^{(+)}_R(E;-\mu^2)\equiv G^{(+)}_0(E)- G_0(-\mu^2)
= \frac{(\mu^2+E)}{(\mu^2+H_0)}G^{(+)}_0(E)
.\label{GR}
\end{equation}
The renormalized T-matrix equation, given above, has a kernel which is 
subtracted at the energy point $-\mu^2$. \ Although the interaction $V$, as 
given by Eq. (\ref{vren}), is not defined for a singular interaction, the
subtracted T-matrix equation is well defined for a regular interaction plus a
delta potential. This justifies {\it a posteriori} the formal
manipulations in obtaing Eq.(\ref{tren3}).
One could instead, work with a finite cut-off in the momentum integrations
and derive the subtracted equation for the T-matrix; then take the
infinite limit  of the cut-off. Of course the subtracted equation for the 
T-matrix is insensible to the cut-off as long the potential is
singular like a delta interaction. 

In the second term of the right-hand side of Eq.(\ref{tren3}), the
propagation over the intermediate states is subtracted at an energy scale
of  $-\mu^2$. The physical information apparently lost in this subtraction
is contained in $T(-\mu^2)$, which in the case of pure contact interaction
is the renormalized coupling constant~\cite{ren}.

If one sets $T(-\mu^2)$ equal to a given operator ($\mu$ independent),
the scheme becomes dependent on the subtraction point, which only
has $\mu$ vanishing as a physical parameter (i.e., the theory is
renormalization group invariant) in the limit $\mu\rightarrow \infty$.
This is a trivial observation when such operator is a regular potential.
Thus,  setting the potential as equal to the T-matrix at $E=-\mu^2$, the
present scheme implies, that the physical T-matrix is found in the limit
$\mu\rightarrow \infty$. 
We will show that the procedure can be applied  for a singular potential 
like OPEP plus contact term, and the limit $\mu\rightarrow \infty$  gives  
finite physical observables with vanishing dependence on $\mu$.
The numerical results obtained by constraining $T(-\mu^2)$ to be
equal to OPEP plus contact term for several values of $\mu$ show
quantitatively the limiting process, while the scattering lengths are 
kept fixed. In this way, one can also verify how good is a scale as low as
the $\rho$-meson mass, in order to approach the limiting
values for the  observables. \ 
Anticipating our numerical findings, the low-energy phase-shifts for the
$^1S_0$, $^3S_1$ and $^3D_1$ channels give a reasonable support to such
low value of $\mu$. \ However, this is not true for the $\epsilon_1$
mixing-parameter, even for energies as low as 20 MeV.

Let us apply the subtracted T-matrix equation for the pure contact 
interaction, and re-obtain some well-known results just for the sake 
of completeness.
In this case, the matrix elements of the T-matrix are independent
of the momentum and, at $E=-\mu^2$, it is given by the renormalized coupling
constant:
\begin{eqnarray}
\langle\vec {p'}|{T}(-\mu^2)|\vec p \rangle=
\frac{1}{2\pi^2}\lambda_{\cal R}(\mu) \ .
\label{lambda1}
\end{eqnarray} 
The renormalized T-matrix in the subtraction point is  given
by $\lambda_{\cal R}(\mu)$, which is the physical information
supplied by the two-body system in order to obtain other observables.
For example, $\lambda_{\cal R}(0)$  
is identified with the scattering length ($a$),
$\lambda_{\cal R}(0) = a$. Solving Eq.(\ref{tren3}), in the 
usual three-dimensional space with condition (\ref{lambda1}), 
\begin{eqnarray}
\langle\vec {p'}|{T}(E)|\vec p \rangle
=\left[{ \frac{2\pi^2}{\lambda_{\cal R}(\mu)}
-\int d^3q\left( \frac 1{E+i\varepsilon-q^2}+
\frac 1{\mu^2+q^2}\right)}\right]^{-1}
\ .
\label{tregd5}
\end{eqnarray}
The on-shell-momentum is $k=\sqrt{E}$ and 
\begin{eqnarray}
\langle\vec {p'}|{T}(E)|\vec p \rangle
=\frac{1}{2\pi^2}\left[{ \frac{1}{\lambda_{\cal R}(\mu)} +
(\mu + i k) }\right]^{-1} \ .
\label{tregd6}
\end{eqnarray} 

In general the matrix elements of a two-body interaction
will depend on the momentum, spin and isospin, such that it will 
be usefull to define our notation for the partial wave 
decomposition. The decomposition of the plane-wave is given by
\begin{eqnarray}
|\vec p ; s m_s;I\rangle =\sqrt{\frac{2}{\pi}
}
\sum_{lsjm_j;I}|p; ls;jm_j\rangle|I\rangle 
\left[{\cal Y}_{ls}^{jm_j}(\hat p)\right]^\dag 
| s m_s\rangle,
\label{pwave}
\end{eqnarray} 
where the quantum numbers $j,m_j,l,s,I$ refer, respectively, to 
the two-body operators: total angular momentum ${\bf J}$, its 
z-component $J_z$, orbital angular momentum ${\bf L}$, total 
spin ${\bf S}$ and total isospin ${\bf I}$.
${\cal Y}_{ls}^{jm_j}(\hat p)$ are the usual orthonormalized 
functions for the coupled angular (${\bf L}$) and spin (${\bf S}$) 
momentum to the total momentum (${\bf J = L+S}$). $\hat p$ defines
the two-dimensional angular variables related to the vector $\vec p$.


The partial wave decomposition of OPEP (\ref{opep}) and,
for simplicity, dropping the unnecessary indices for the total 
angular momentum and spin in $V_\pi$, is given by 
\begin{eqnarray}
V^{(l^\prime l)}_{\pi , \alpha }(p^\prime,p)
&=&-\frac{g_a^2[2 I_\alpha(I_\alpha+1)-3]}{(8\pi)^2f_\pi^2} 
\int d\hat p \int d\hat p^\prime
\left({\cal Y}_{l\frac 12}^{jm_j}(\hat p)\right)^\dag
\times \nonumber\\ &&
\left[\frac{\vec\sigma_1\cdot(\vec {p'}-\vec {p})  
 \vec\sigma_2\cdot(\vec {p'}-\vec {p})}
{ (\vec{p'}-\vec{p})^2+m_\pi^2}\right] 
\left({\cal Y}_{l^\prime \frac 12}^{jm_j}(\hat p^\prime)\right)
\ , \label{opep3}
\end{eqnarray} 
where for the singlet state $\alpha = s$  ($I_s=1$) and
for triplet state $\alpha =t$ ($I_t =0$).

So, in the angular projected equation for the T-matrix, the singlet $^1S_0$
state matrix element for the OPEP, corresponding to eq.~(\ref{opep}), 
is given as
\begin{eqnarray}
V^{(00)}_{\pi , s}(p^\prime,p)
&=&
\frac{g_a^2}{32\pi f_\pi^2}
\left(2-\int^1_{-1}dx \frac{m_\pi^2}{p^2+{p'}^2-2 p {p'}x+m_\pi^2}
\right).
\label{sing}
\end{eqnarray} 
In the limit of $p$ or $p'$ going to infinity the matrix element 
goes to a constant:
\begin{equation}
\lim_{p'\rightarrow \infty}
V^{(00)}_{\pi , s}(p^\prime,p) =\frac{g_a^2}{16\pi f_\pi^2} 
\ . \label{lim1s0}
\end{equation}

The corresponding angular momentum projected matrix elements,
for triplet $^3S_1-^3D_1$  coupled channels, 
($V^{(l'l)}_{\pi , t}$, with $l',l=0, 2$) are:
\begin{eqnarray}
V^{(0,0)}_{\pi ,t}(p^\prime,p)
&=&\frac{g_a^2}{32\pi f_\pi^2}\left(2- \int^1_{-1}dx 
\frac{m_\pi^2}{p^2+{p'}^2-2 p {p'}x+m_\pi^2}\right) \ ,
\label{ss} \\
V^{(20)}_{\pi , t}(p^\prime,p)
&=&\frac{g_a^2\sqrt{2}}{16\pi f_\pi^2}
\int^1_{-1}dx \frac{{p'}^2-2 p p'x +p^2(\frac32 x^2-\frac12)}
{p^2+{p'}^2-2 p {p'}x+m_\pi^2}
, \label{ds} \\
V^{(22)}_{\pi , t}(p^\prime,p)
&=&\frac{g_a^2}{32\pi f_\pi^2}
\int^1_{-1}dx \frac{2 p p'x -(p^2 +p'^2)(\frac32 x^2-\frac12)}
{p^2+{p'}^2-2 p {p'}x+m_\pi^2} \ ,
\label{dd}
\end{eqnarray} 
and $V^{(02)}_{\pi , t}(p', p)=V^{(20)}_{\pi ,t}(p, p')$.

In the limit of $p$ or $p'$ going to infinity, the matrix elements
of  OPEP in the $^3S_1-^3D_1$ channels are zero or constant:
\begin{eqnarray}
\lim_{p'\rightarrow \infty}V^{(00)}_{\pi,t}(p',p)
&=&\frac{g_a^2}{16\pi f_\pi^2} ,\;\;\;\; 
\lim_{p'\rightarrow \infty}V^{(22)}_{\pi,t}(p',p)
=0 , 
\nonumber\\
\lim_{p'\rightarrow \infty}V^{(20)}_{\pi,t}(p', p)
&=&\frac{g_a^2\sqrt{2}}{8\pi f_\pi^2} \ ,\;\;\;\;
\lim_{p\rightarrow \infty}V^{(20)}_{\pi,t}(p',p)
=0 \ ; \label{lim}
\end{eqnarray} 
consequently, just one subtraction is enough to obtain a finite 
T-matrix, once $T(-\mu^2)$ is given. 

The trivial choice for $T(-\mu^2)$ is the OPEP operator added to
a delta interaction. \
For the singlet renormalized T-matrix, with the renormalized strength
of the contact interaction ($\lambda_{{\cal R},s}(\mu)$)
fixed by the singlet scattering length $a_s=-$23.7\ fm, we have
\begin{eqnarray}
T^{(00)}_s(p',p;-\mu^2)=V^{(00)}_{\pi ,s}(p',p) +
\lambda_{{\cal R},s}(\mu) \ ,
\label{tsing}
\end{eqnarray} 
where we have dropped the upper index in $\lambda $ for simplicity.
As for the coupled $^3S_1-^3D_1$ channels, the renormalized T-matrix 
for each channel ($l, l' =0,2$) is given by
\begin{eqnarray}
T^{(l'l)}_t(p',p;-\mu^2)=V^{ (l'l)}_{\pi ,t}(p',p) +
\lambda_{{\cal R},t}(\mu) \delta_{l'0} \delta_{l0}\ .
\label{ttrip}
\end{eqnarray} 
In this case, the renormalized strength of the contact interaction 
($\lambda_{{\cal R},t}(\mu)$) is fixed by the triplet scattering 
length, $a_t=5.4\ $ fm.

Next, the formalism in momentum space, for the renormalized 
partial-wave T-matrix, using one subtraction in the propagator at
an energy $-\mu^2$ is concluded by writting the
integral equations  corresponding to Eq. (\ref{tr1}) 
where one should replace the matrix elements for $T(-\mu^2)$ by the 
corresponding above expressions [for the singlet state, eq. (\ref{tsing}); 
for the triplet state, eq. (\ref{ttrip})]. \ 
For the singlet state, we have
\begin{eqnarray}
T^{(00)}_s(p',p;k^2) &=&  T^{(00)}_s(p',p;-\mu^2) + 
\nonumber\\ &&
\frac{2}{\pi}\int_0^\infty dq q^2
\left(\frac{\mu^2+k^2}{\mu^2+q^2}\right)
\frac{ T^{(00)}_s(p',q;-\mu^2)}{k^2-q^2+i\epsilon} 
T^{(00)}_s(p',p; k^2)
\label{tsing2};
\end{eqnarray} 
and for the triplet state, with $l_i =$ 0, 2 $(i=1,2,3)$, 
the corresponding coupled equation is given by:
\begin{eqnarray}
T^{(l_1l_2)}_t(p',p;k^2)&=& 
T^{(l_1l_2)}_t(p',p;-\mu^2) + 
\nonumber\\ &&
\frac{2}{\pi}\sum_{l_3}\int_0^\infty dq q^2
\left(\frac{\mu^2+k^2}{\mu^2+q^2}\right)
\frac{ T^{(l_1l_3)}_t(p',q;-\mu^2)}{k^2-q^2+i\epsilon} 
T^{(l_3l_2)}_t(p',p;k^2)
\label{ttrip2}
\end{eqnarray} 
These eqs. (\ref{tsing2}) and (\ref{ttrip2}) for a regular potential
clearly converge to the exact
original T-matrix in the limit $\mu^2\to\infty$, with the corresponding
Born terms given by the potential.

\section{Numerical results and conclusions}
In the following, we present the numerical results for the 
observables, in the $^1S_0$ and $^3S_1-^3D_1$ channels, calculated
from the leading order NN interaction, with the subtracted 
eq.~(\ref{tren3}) and the T-matrix elements 
at scale $\mu$ defined by eqs.~(\ref{tsing}) and (\ref{ttrip}).
Distinct numerical procedures can be used to obtain the final results, from
the equations (\ref{tsing2}) and (\ref{ttrip2}).
In our numerical calculation, we have followed the one described 
in Ref. (\cite{at}),
where the T-matrix is related to an auxiliary real matrix, which is solved
by the inverse matrix approach.
Considered that in the present work we are not interested in high 
numerical accuracy, 
any of the usual numerical procedures should give 
about the same results we obtain, once a large enough number of points is taken
in the numerical discretization of the momentum space. \ In our calculation, 
when discretizing the momentum space ($p$) from 0 to $\infty$, we
have used a Gaussian mesh with up to 100 points, mapping the interval
$[-1,1]$ to $[0,\infty]$ through the expression 
$ p = p_0 (1+x)/(1-x) .$
$p_0$ is adjusted to optimize the convergence of the results, and in 
practice should be comparable with $\mu$, when a minimum number of mesh points 
is required.

In Table I, for several values of $\mu$, we show the corresponding results 
for the  singlet and triplet effective ranges, and also for deuteron 
observables (binding energies and ratio $\eta_D$). 
The ratio $\eta_D$ is given by
\begin{equation}
\eta_D= \lim_{k\to i k_B}\frac{T^{(02)}_t(k,k;k^2)}{T^{(00)}_t(k,k;k^2)} 
\label{eta},
\end{equation}
where the deuteron binding energy is $E=-k^2_B$.
The effective ranges, for the singlet state ($r_{0,s}$) and for the triplet
state ($r_{0,t}$) and the corresponding scattering lengths ($a_s$ and $a_t$) 
are related to the $s-$wave phase shifts, according to the effective range 
expansion:
\begin{eqnarray}
&&k\cot\left(\delta_{0,\alpha}^{(\alpha)}(k)\right)=-\frac{1}{a_\alpha}
+\frac{1}{2}r_{0,\alpha} k^2 + {\cal O}(k^4) \ .
\label{eff}
\end{eqnarray}

\vskip 0.5cm
{\small {\bf Table I}
\ \ Low energy  n-p and deuteron observables compared with
data. Singlet ($r_{0,s}$) and triplet ($r_{0,t}$) effective ranges,
deuteron binding energy ($B_D$) and ratio $\eta_D$ are given for 
several values of the single parameter $\mu$ which is used in 
our renormalization scheme. The $\lambda_{{\cal R},s}(\mu)$ and
$\lambda_{{\cal R},t}(\mu)$ are the strenghts of the $\delta-$interactions
which were added to the OPEP and adjusted to the corresponding
scattering lengths, $a_s=$-23.7 fm and $a_t=$ 5.4 fm.
The results were compared with data obtained from Ref.~\cite{nij}.}
\begin{center}
\begin{tabular}{ccccccc} 
\hline \hline
$\mu \ ($fm$^{-1})$ 
& $\lambda_{{\cal{R}},s}$(fm) 
& $r_{0,s}$(fm) 
& $\lambda_{{\cal{R}},t}$(fm) 
& $r_{0,t}$(fm) 
& $B_D$  (MeV) 
& $\eta_D$   \\ \hline 
   4  & -0.8806 & 1.332 & -0.2281 & 1.364 & 1.977  & 0.02808 \\
   10 & -0.7570 & 1.345 &  21.741 & 1.536 & 2.084  & 0.02904 \\
   30 & -0.6977 & 1.347 & -0.3776 & 1.582 & 2.114  & 0.02933 \\
Ref.\cite{nij}&-& 2.73 & -      & 1.75  & 2.2246 & 0.0256  \\ 
\hline \hline
\end{tabular} 
\end{center}
\vskip 1cm

\ The results in  Table I are compared with experimental results obtained from
Ref.\cite{nij}, and show a good qualitative agreement, considering that we are 
only using the leading order term of the effective NN interaction.
As it is shown, the agreement improves as we increase the value
of $\mu$. \ The results in the $^1S_0$ channel could be improved by 
adding higher-order terms in the effective interaction, 
as already one can deduce from previous approaches used by other authors
in EFT~\cite{kaplan,rho,mehen}. \ It is also 
important to observe that for $\mu > 4$fm$^{-1}$, $r_{0,s}$ is
pratically stable, such that it is insensible to 
nucleon propagation for higher momenta once the scattering length is fixed.
The insensibility to $\mu > 4$fm$^{-1}$ is also found in the singlet 
phase-shifts up to laboratory energies of about 100 MeV.

In the triplet channel, we observe 
a large variation of $\lambda_{{\cal R},t}(\mu)$ as the scale is moved
to higher values. The oscillating behaviour is a consequence of the
zeroes of $\lambda_{{\cal R},t}(\mu)$  at those scales where
only OPEP is enough to reproduce the triplet scattering length.
However, in spite of such variation, the triplet low-energy and deuteron 
observables are quite stable, and converging in the limit of $\mu $ 
going to infinity. From the observables we have examined, as we pointed
out in our discussion, the slowest rate of convergence is found for the
mixing parameter $\epsilon_1$, which is defined through the on-shell S$-$ 
and T$-$matrices in the coupled channel. With an implicit energy dependence, 
such relation can be written as a two-dimensional matrix and 
parametrized according to Ref.~\cite{stapp}: 
\vskip 0.5cm
\begin{eqnarray}
S=1-2 i k T_t
=\left(\begin{array}{cc}
\cos (2\epsilon_1) e^{2i\delta_{0,t}} &
i\sin (2\epsilon_1) e^{i(\delta_{0,t}+\delta_{2,t})} \\
i\sin (2\epsilon_1) e^{i(\delta_{0,t}+\delta_{2,t})} &
\cos (2\epsilon_1) e^{2i\delta_{2,t}}
\end{array}
\right) .
\label{mixing}
\end{eqnarray} 

In Figs. 1, 2 and 3  we show the results for the neutron-proton phase-shifts
$^1S_0$, $^3S_1$ and $^3D_1$, respectively. \ The results shown in Fig. 1 
are pratically insensible to nucleon-nucleon propagation for momenta higher 
than 4 fm$^{-1}$, in agreement with 
results shown for the singlet effective range (Table I), such that no
improvement can be expected for the singlet observables with just one
subtraction/parameter. \ In Fig. 2, the good qualitative fit to data 
obtained with the scaling parameter $\mu=4$ fm $^{-1}$ is improved 
when taking larger value for $\mu$.
The subtraction point above $\mu > 4 {\rm fm}^{-1}$ has some effect,
as also anticipated from the values of scattering length and effective
range (see Table I); but the effect is not strong above 30 fm$^{-1}$. 
In Fig. 3, we observe that the insensibility to nucleon-nucleon 
propagation at higher momenta is also found in the $D-$wave. In this state
the phase-shift is clearly dominated by the OPEP, as also shown
in ref.~\cite{BR}, in a first order perturbative calculation.

In Fig. 4 we present our results for the mixing parameter, $\epsilon_1$,
which is obtained using eq.~(\ref{mixing}). \ 
The results are insentive to $\mu > 4 {\rm fm}^{-1}$ only 
for $E_{lab} <$ 10 MeV, and show a fair reproduction of the Nijmegen values. 
This follows due to the dominance of  OPEP 
in  the slope of $\epsilon_1$ at zero energy\cite{mix}. \ Above 10 MeV,
the mixing parameter  shows a dependence on the values of $\mu$, which is 
reduced by increasing $\mu$ above 30${\rm fm}^{-1}$, while the agreement 
with data is improved as $\mu\to\infty$. The mixing parameter is the most
sensitive parameter in respect to changes in the subtraction point for 
$E_{lab}\ >$ 10 MeV and for $\mu$ between 4 and 30 fm$^{-1}$. Our results 
in the limit of $\mu\rightarrow \infty$ are consistent with the ones 
obtained by Ballot and Robillota \cite{BR}, where they have used a cut-off
parameter. 

The results presented in Table I and Figs. 1-4, when compared with
fitted data~\cite{nij} and with other  calculations with more fitting
parameters \cite{bira,kaplan,cohen}, are a clear evidence of the dominance
of the leading order term (OPEP) in low-energy NN observables. This
implies that one should first renormalize the T-matrix for the leading
order term and then add higher order terms to the NN interaction in
the chiral expansion, or in the EFT expansion.  
Our results for the phase-shifts and mixing parameter
show that the mixing parameter is the most sensitive
observable to the subtraction point used in the renormalization procedure. The 
phase-shifts for  $^1S_0$ state and for the $^3D_1$ are pratically 
insensitive to $\mu$ larger than 4 fm$^{-1}$, with an evidence that
$^3D_1$ is dominated by the OPEP, in agreement with previous 
calculations\cite{BR}. \ The present results confirm that the addition of
higher order singular terms in the potential are needed in order to fit
the singlet effective range and corresponding phase-shifts. 
This observation was already verified in several EFT
approaches~\cite{bira,kaplan,meissner,birse,rho,cohen,gegelia}, 
without however addressing to the full predictibility of the
leading order term. \ This lead some authors to affect negatively the nice
fitting obtained for the triplet channel when trying to improve the
observables for the singlet channel~\cite{kaplan}.
In order to avoid this problem, it is important to parametrize the higher
order corrections, such that it affects mostly the singlet channel (where
the leading order fails to give a good description also to the effective
range, independently of the renormalization point).
In the $^3S_1-^3D_1$ channels, the higher order terms will become more 
important for laboratory energies above 50 MeV, where the leading order
results deviate from the data (as shown in Figs. 2 and 4). 

Considering the results obtained and shown in Figs. 1-4, we would like
to make some comments about the renormalization scale
$\mu$, which appears in our approach as the 
point where we introduce a subtraction in the nucleon-nucleon propagator. 
(This parameter is directly related to the ``cut-off" parameter of other
approaches.)
Weinberg~\cite{wei} has originally suggested to cut the virtual
propagation of the NN system at the scale $m_\rho$, which implies
$\mu \sim $ 4 fm$^{-1}$, and this suggestion is consistent with our
results, as it is almost exactly verified for the $^1S_0$ and 
$^3D_1$ phase shifts (see Figs. 1 and 3), and also approximately for the
$^3S_1$ phase shifts (Fig. 2). \ We further observe that our results
for $\mu\to\infty$ (numerically, we have used $\mu\le$ 60 fm$^{-1}$)
are about the same we got for the $^1S_0$ and $^3D_1$, and improves
the results we obtain for the $^3S_1$ case, for the mixing
parameter, and also related observables, as shown in Table I.
Such results, particularly the results for the mixing parameter, lead us
to conclude that Weinberg's suggestion can only be understood as 
a qualitative statement, and as such, it is verified. However,
we observe that 
the mixing parameter, which is very sensible to this renormalization scale, 
is violating this statement even for $E_{lab}>$ 20 MeV. This result implies 
that some essential physical information for such observable is being lost when
using $\mu\sim m_\rho$, and as we observe for other observables, there is
no danger in increasing the value of $\mu$. \ Another important observation, 
related to a statement of caution made in ref.\cite{rho}, is that we are not 
{\it unnecessarily complicating the theory}, or {\it introducing irrelevant 
degrees of freedom} when considering $\mu\to\infty$, as it is possible to 
occur in other EFT approaches.  \ 
Our conclusion for a practical calculation with
the leading order term is that one should use a momentum renormalization
parameter of the order of $\sim 4$ fm$^{-1}$ (as suggested in \cite{wei})
to obtain qualitative results for the low energy NN observables. But,
to improve the results for some observables, like the triplet mixing parameter 
shown in Fig. 4 and also the $^3S_1$ phase shifts, there is no danger in 
extending
$\mu\to\infty$, as the previous results are not affected by increasing such
value for the scale.

In conclusion, we have examined 
the complete renormalization of the leading order term of the neutron-proton
interaction. We suggest the use of a T-matrix with a subtracted propagator,
at an energy $-\mu^2$, and we have concluded 
about an overall insensitivity of the low-energy singlet and triplet
phase-shifts for momentum scale $\mu$ larger than 4 fm$^{-1}$. 
Such result supports the qualitative suggestion presented in Ref.~\cite{wei}. 
\ However, the results for the mixing parameter 
(see Fig. 4) show that this observable is more sensible to the
renormalization scale, and therefore more appropriate than the diagonal 
phase-shift results to be described by an EFT. \ Our results for this
observable show that $\mu\to\infty$ is necessary to have a better 
agreement with data. And notice that a larger value of $\mu$
is also consistent with the previous results obtained for the 
phase-shifts. 
We can conclude that, in our renormalization scheme, $\mu$ vanishes as a
physical parameter, and the theory is renormalization group invariant
in the limit $\mu\rightarrow \infty$ ~\footnote{By excluding the
non-diagonal matrix element one could reach a wrong conclusion about $\mu$.}. 

It has been achieved with this work that the renormalized T-matrix  for
the leading order interaction has a well defined limit for the physical
observables for $\mu \to \infty$ with the constraint that $T(-\mu^2)$
being equal to OPEP plus contact interaction, which also
is supported by the data in the triplet channel.
The following step to be investigated is the 
inclusion of the next order term in the EFT expansion of 
the NN interaction, i.e., the Laplacian of the delta potential.
In this case, more subtractions are required to construct the
renormalized T-matrix, starting from the T-matrix (for the leading order)
obtained in this work.

We thank Drs. A. Delfino, M. Malheiro and P. Sauer for useful discussions.
This work was partially supported by Funda\c c\~ao de Amparo \`a Pesquisa do  
Estado de S\~ao Paulo (FAPESP) and Conselho Nacional de Desenvolvimento
Cient\'\i fico e Tecnol\'ogico (CNPq).

\vspace{-0.5cm} 

 
\newpage
\begin{figure}[tbp]
\vspace*{0.5cm} \epsfxsize=15cm
\epsfysize=15cm \centerline{\epsfbox{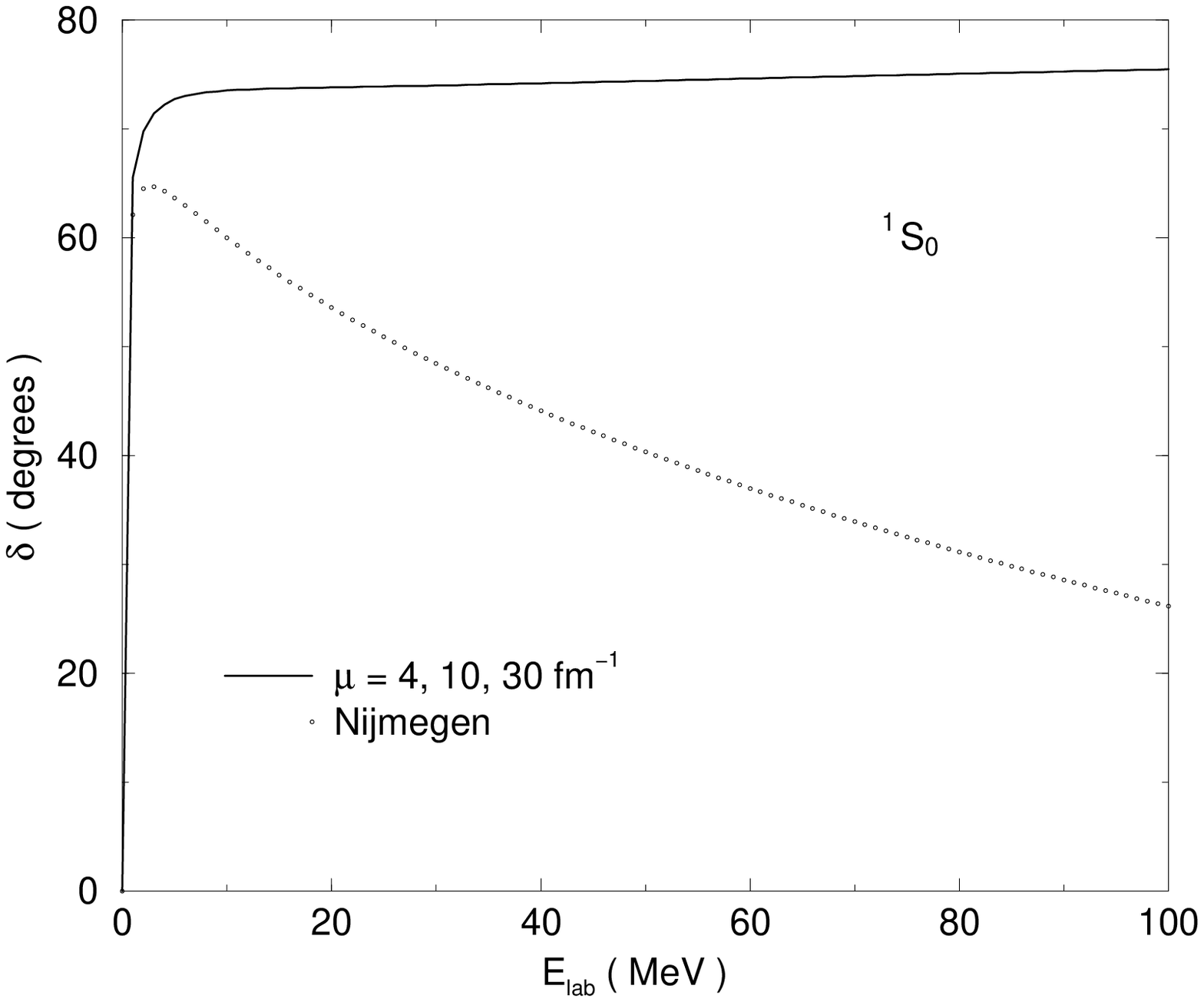}}
\caption[dummy0]
{Neutron-Proton phase shifts for the $^1S_0$ state, $\delta_{0,s}$.
The solid line is the model result, for different parameters $\mu$; 
the dotted line gives the fitting to data from Ref.~\cite{nij}.}
\end{figure}
\newpage
\begin{figure}[tbp]
\vspace*{0.5cm} \epsfxsize=15cm
\epsfysize=15cm \centerline{\epsfbox{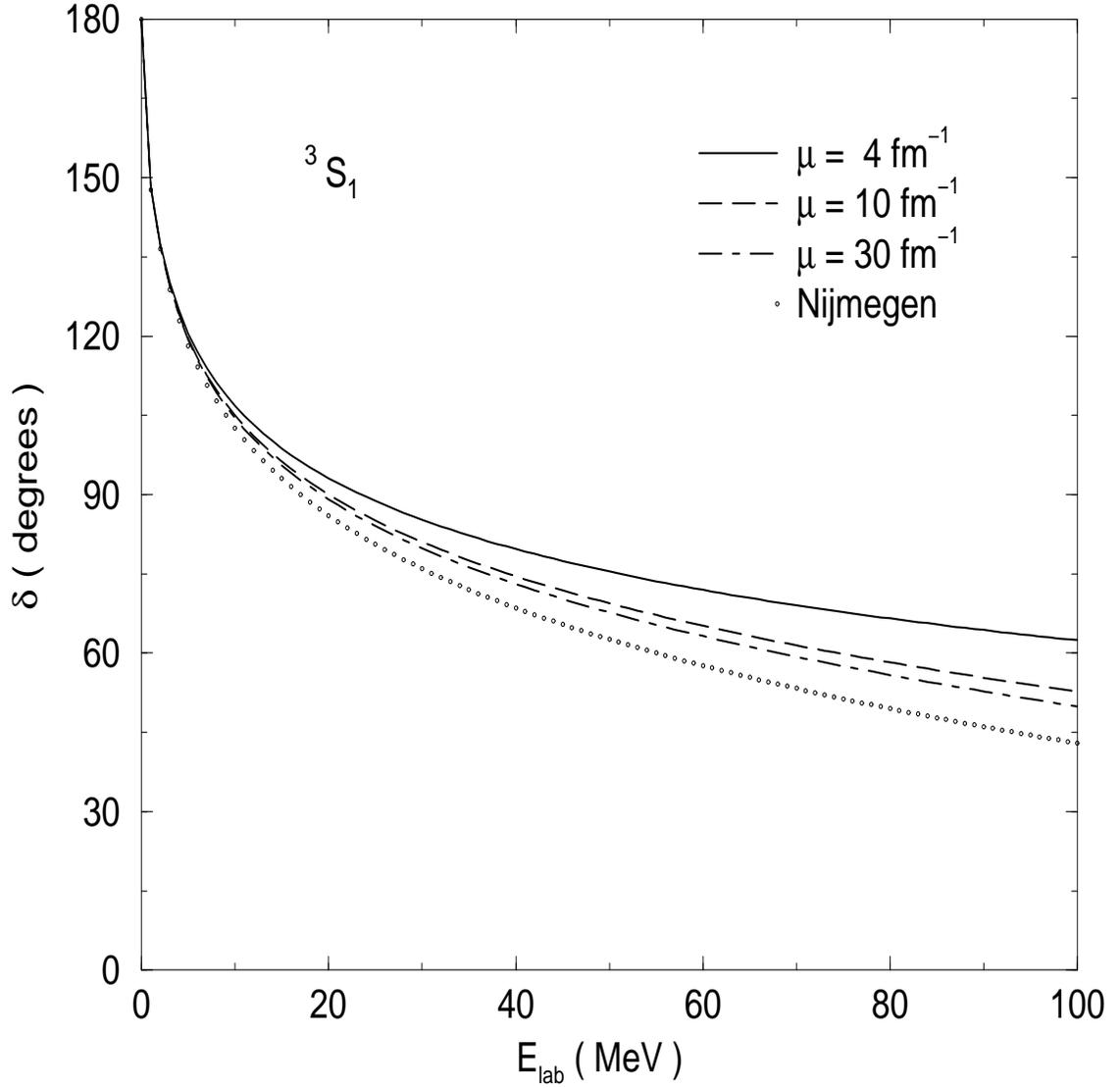}}
\caption[dummy0]
{Neutron-Proton phase shifts for the $^3S_1$ state,
$\delta_{0,t}$.
Calculations are presented for $\mu=$ 4, 10 and 30 fm$^{-1}$; 
the dotted line gives the fitting to data from Ref.~\cite{nij}.}
\end{figure}
\newpage
\begin{figure}[tbp]
\vspace*{0.5cm} \epsfxsize=15cm
\epsfysize=15cm \centerline{\epsfbox{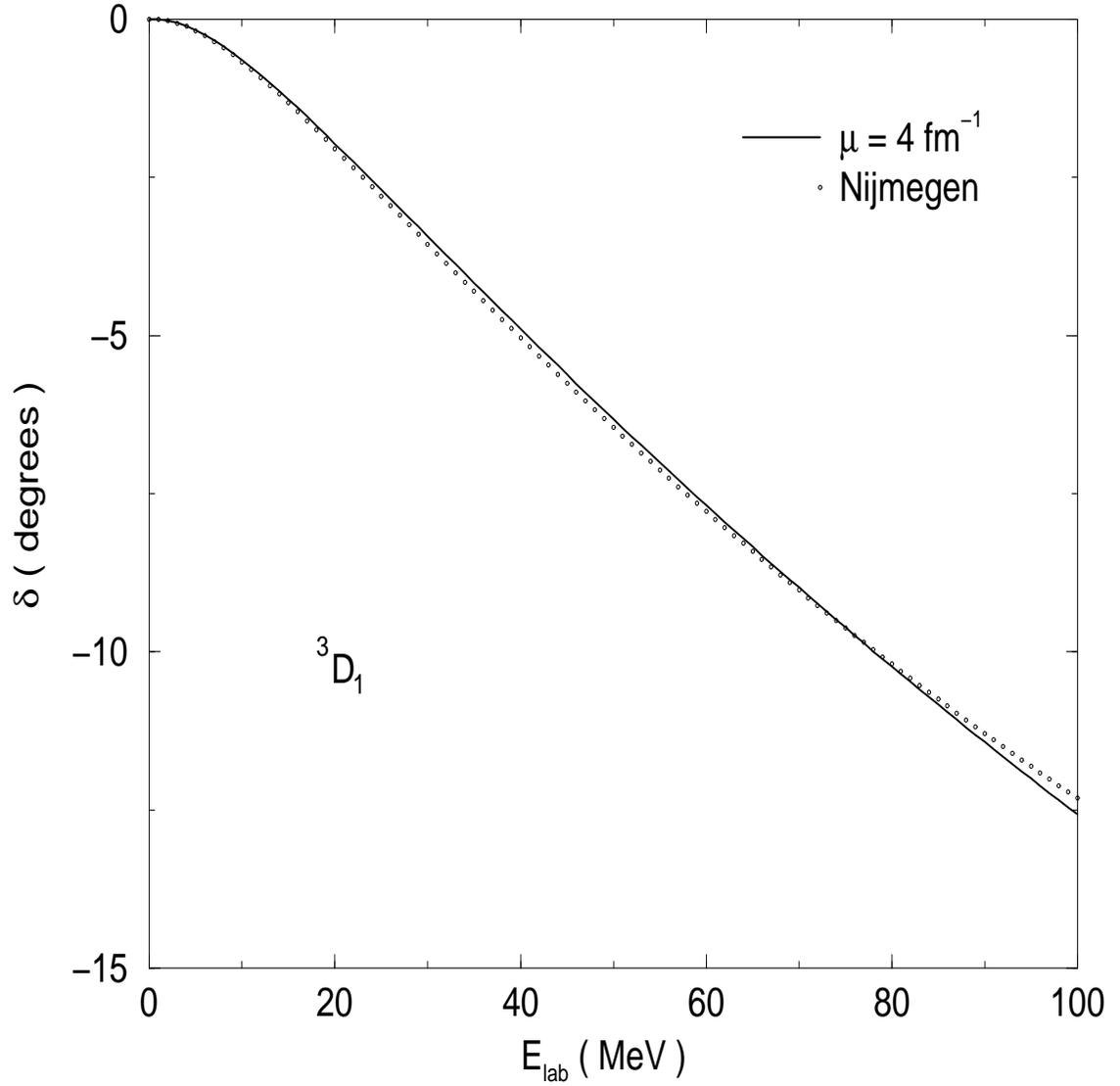}}
\caption[dummy0]
{Neutron-Proton phase shifts for $^3D_1$ state, $\delta_{2,t}$. The 
solid line is the present calculation, for different parameters $\mu\ge $4 
fm$^{-1}$; the dotted line gives the fitting to data from
Ref.~\cite{nij}.} 
\end{figure}
\newpage
\begin{figure}[tbp]
\vspace*{0.5cm} \epsfxsize=15cm
\epsfysize=15cm \centerline{\epsfbox{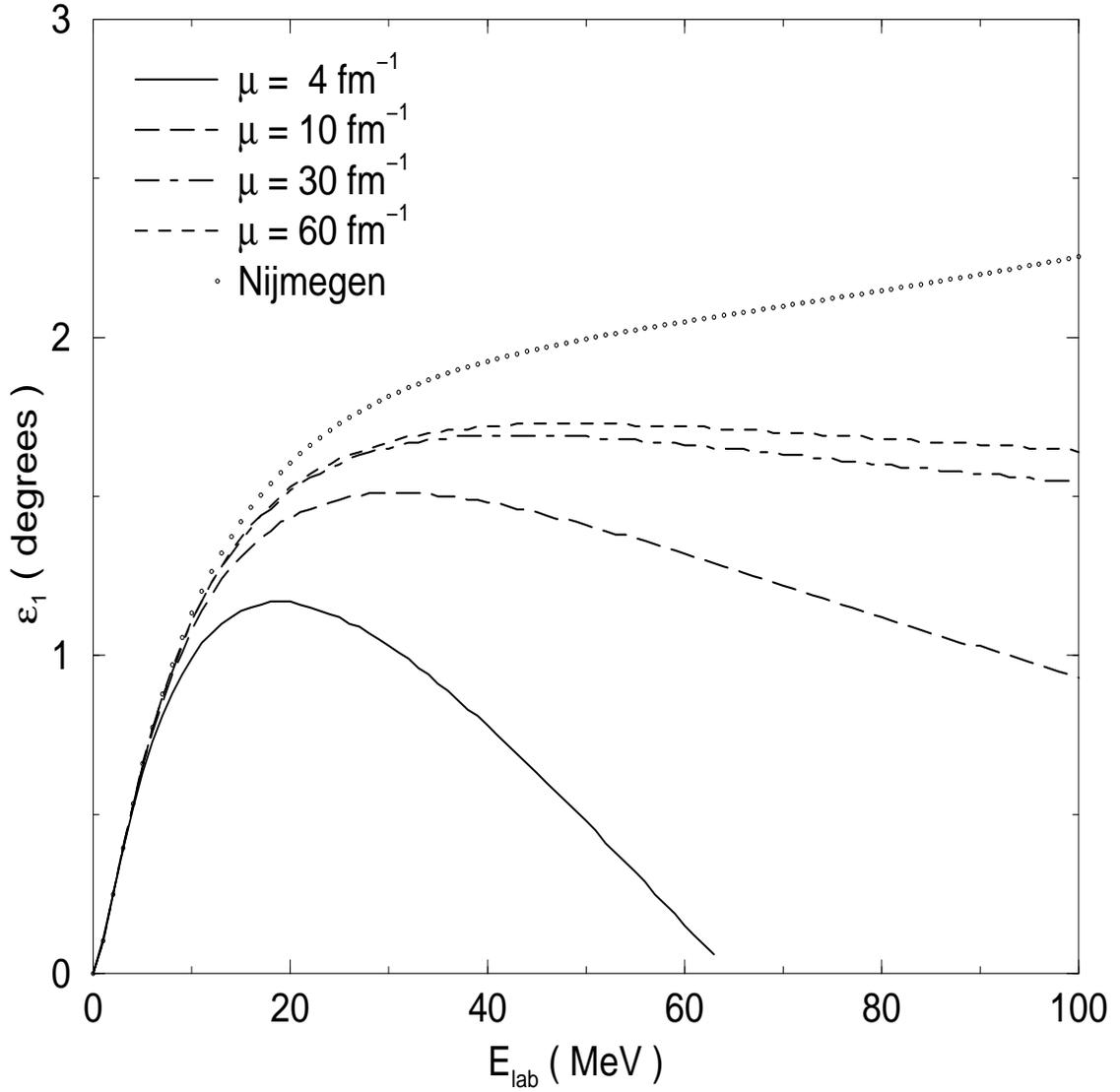}}
\caption[dummy0]
{Results for the mixing parameter (in degrees) given by 
eq.~(\ref{mixing}), for $\mu=$ 4, 10, 30 and 60 fm$^{-1}$. 
The dotted line gives the fitting to data from Ref.~\cite{nij}.}
\end{figure}
\end{document}